\documentclass[journal, 10pt]{IEEEtran}
\IEEEoverridecommandlockouts

\usepackage{amsmath}
\usepackage{mathtools}
\usepackage{cuted}
\usepackage{revsymb}
\usepackage[tight,footnotesize]{subfigure}
\usepackage{balance}
\usepackage{xcolor}
\usepackage{eqnarray}
\usepackage{multicol}
\usepackage{commath}
\usepackage{graphicx}
\usepackage{textcomp}
\DeclareGraphicsExtensions{.eps}
\graphicspath{{./Figures/}}
\usepackage{multicol}
\usepackage{algorithm,algorithmic}
\usepackage{subfiles}
\usepackage{mathcomp}
\usepackage{subfig}
\usepackage{cite}
\usepackage{cases}
\usepackage{amssymb}

\usepackage{verbatim}
\normalsize

\usepackage[normalem]{ulem}

\begin{document}

\title{A Thermal  Study of Terahertz Induced Protein Interactions}

 \author{Hadeel Elayan, Samar Elmaadawy, Andrew W. Eckford, Raviraj Adve and Josep Jornet 
%

\thanks{We would like to acknowledge the support of the National Science and Engineering Research Council, Canada, through its Discovery Grant program.}
\thanks{H. Elayan and R. Adve are with the Edward S. Rogers Department of Electrical and Computer Engineering, University of Toronto, Ontario, Canada, M5S 3G4. (Corresponding author: Hadeel Elayan; email: hadeel.mohammad@mail.utoronto.ca.)
}
\thanks{S. Elmaadawy and J. Jornet  are with the Department of Electrical and Computer
Engineering, Northeastern University, Boston, MA 02115 USA.}
\thanks{ A. Eckford is  with the Department of Electrical Engineering and Computer Science, York University, Ontario, Canada, M3J 1P3.}}

\maketitle

\begin{abstract} Proteins can be regarded as thermal nanosensors in an intra-body network. Upon being stimulated by Terahertz (THz) frequencies that match their vibrational modes, protein molecules experience resonant absorption and dissipate their energy as heat, undergoing a thermal process. This paper aims to analyze the effect of THz signaling on the protein heat dissipation mechanism. We therefore deploy a mathematical framework based on the heat diffusion model to characterize how proteins absorb THz-electromagnetic (EM) energy from the stimulating EM fields and subsequently release this energy as heat to their immediate surroundings. We also conduct a parametric study to explain the impact of the signal power, pulse duration, and inter-particle distance on the protein thermal analysis. In addition, we demonstrate the relationship between the change in temperature and the opening probability of thermally-gated ion channels. Our results indicate that a controlled temperature change can be achieved in an intra-body environment by exciting protein particles at their resonant frequencies. We further verify our results numerically using COMSOL Multiphysics\textsuperscript{\textregistered} and introduce an experimental framework that assesses the effects of THz radiation on protein particles. We conclude that under controlled heating, protein molecules can serve as hotspots that impact thermally-gated ion channels. 
Through the presented work, we infer that the heating process can be engineered on different time and length scales by controlling the THz-EM signal input.
\end{abstract}

%

 \section{Introduction}

 The Internet of Bio-Nano Things (IoBNT)  has been introduced as a paradigm-shifting concept for communication and network engineering~\cite{akyildiz2015internet}.  It refers to the interconnectivity of various biological and artificial nanomachines, which can be connected to the Internet and to each other, creating a network of integrated, intelligent, and self-organizing systems. Sensing and actuation are key components of the IoBNT. Sensors are used to collect data on various biological parameters as well as analyze circulating biomarkers
in body fluids for the diagnosis
of diseases, including different
types of cancer. Actuators, on the other hand, are used to control the release of drugs and to activate
or inhibit the response of neurons,
in the context of neurodevelopmental
and neurodegenerative diseases~\cite{10101695}. Therefore, the use of the IoBNT has the potential to revolutionize healthcare by enabling personalized medicine, early detection of diseases, and targeted treatments~\cite{lemic2021survey}.

 Researchers  started investigating high frequency electromagnetic (EM) waves to attain miniaturized sensors with the desired resolution and sensitivity~\cite{lee2015highly}. Specifically, the terahertz  (THz) frequency band, defined between 0.1 and 10 THz, has drawn attention as a key enabler of communication at the nanoscale. As such, devices with ultra-small footprints, plasmonic waveguides, as well as nanoantennas~\cite{yahiaoui2016terahertz,nafari2017modeling} have been developed. With the emergence of these devices, a substantial amount of work has been conducted to characterize and model THz propagation through various body tissues~\cite{elayan2017terahertz,elayan2017multi,elayan2018end}.  In addition, a hybrid THz-molecular model that capitalizes on stimulating proteins in the human body using THz signals has been introduced. In the presented system, the transduced energy directly interfaces with either natural or engineered proteins to invoke controlled intra-body interactions~\cite{elayan2020information}.  

 Among the biological entities, proteins perform a wide range of functions, including catalyzing chemical reactions, transporting molecules across cell membranes, and transmitting signals within and between cells~\cite{kessel2010introduction}. These functions are achieved by the precise arrangement of amino acids into a specific three-dimensional shape. Importantly, for our purposes, nanoscale devices and sensors employed in an IoBNT network can be tailored to interact with proteins. For instance, nanosensors could be engineered to detect specific proteins or monitor changes in their concentrations. Conversely, proteins can interact with nanomaterials, such as nanoparticles or nanotubes, commonly utilized in IoBNT devices, resulting in phenomena like protein adsorption. Understanding how proteins adhere to nanomaterial surfaces is pivotal for various applications, including biosensing, drug delivery, and biomedical implants. IoBNT platforms can utilize this comprehension to develop more effective and efficient nanodevices for healthcare, environmental monitoring, and other fields where nanotechnology plays a crucial role.

 Proteins also exhibit collective vibrational modes that couple very well with frequencies in the THz frequency band~\cite{turton2014terahertz}. These modes provide information about protein conformational changes, ligand binding and oxidation 
 states~\cite{xu2006probing}. The aforementioned interactions can therefore trigger unique changes in vibration that can be used in detection and diagnoses~\cite{9819908}. Despite the potential associated with THz-EM nanonetworks, the thermal impact of THz waves in an intra-body environment is still an ongoing research area. In fact, the latest IEEE Standard for Safety Levels (IEEE C95.1)  with respect to human exposure to EM fields has been limited to frequencies up to 300 GHz (whole body average)~\cite{8859679}. As such, heat transfer  procedures in the THz frequency band require  a deep understanding of the basic mechanisms of heat transport along with adequate mathematical modeling. The source power density, the relevant time scale of the deposited energy as well as the differences in boundary and initial conditions strongly impact the energy balance~\cite{hristov2019bio}.

 Studies of the photothermal effect of THz waves have been limited to the cellular~\cite{elayan2017photothermal} and tissue level~\cite{reddy2022multi}, where  both cases  assume a homogeneous medium of identical particles. However, homogeneous models neglect possible field localizations and local temperature build up at smaller, inhomogeneous scales~\cite{saviz2013theoretical}. Therefore, the exact mechanisms by which THz radiation affects protein molecules is not yet fully understood.
 
 By extending the precision of the thermal effects below cellular dimensions using proteins, new fields of applications for THz signals in medicine and biology may arise. In this work, we model the thermal effects of  THz stimulated proteins using the Goldenberg and Tranter heat transfer model~\cite{goldenberg1952heat}. This model provides the transient solution of the heat conduction problem in which a
sphere generating heat is placed in a medium of different thermal properties. We explain the resonant absorption phenomenon experienced by proteins when triggered by a frequency that matches the frequency of one of its vibrational modes. Accordingly, protein molecules act as thermal nanosensors in an intra-body network and can play a vital role in triggering membrane receptors. We show that apart from light, THz-EM waves serve as an alternative solution to modulate protein activities by inducing controlled temperature changes. From this perspective, we make the following  contributions:
\begin{itemize} 

\item We  calculate the temperature change of protein particles induced by THz-EM waves using the  Goldenberg and Tranter heat transfer model~\cite{goldenberg1952heat}. 
   
\item We analyze the temperature change as a function of the transmitter (Tx) power and the duration of irradiation. 

\item We explain the relationship between temperature and the opening probability of thermally-gated ion channels.

\item We present numerical simulations via COMSOL Multiphysics\textsuperscript{\textregistered} to complement the analytical framework.

\item We propose an experimental framework to present a proof-of-concept in regards of the impact of THz radiation on 
protein particles. 

\end{itemize}

 The rest of the paper is organized as follows. In Sec.~\ref{Sec:Sec2}, we illustrate the physical basis and the mathematical model of the THz induced protein particles. In Sec.~\ref{Sec:Sec3}, we present the heat transfer model based on the Goldenberg and Tranter analysis. In Sec.~\ref{Sec:Sec4}, we discuss a number of temperature-gated ion channels and the impact of temperature on their kinetic reactions. In Sec.~\ref{Sec:Sec5}, we  demonstrate our results based on the developed models. Finally, we draw our conclusions in Sec.~\ref{Sec:Sec6}.

\section{System Model}
\label{Sec:Sec2}

\subsection{Physical Basis}

Molecular absorption refers to the process by which atoms in the molecule absorbs EM radiation due to the excitation of its electrons to higher energy levels. The energy of the absorbed EM wave must match the energy difference between the ground state and the excited state of the molecule. While molecular absorption can occur at any frequency at which a molecule has an electronic transition, \textit{resonant absorption} is typically limited to specific frequencies. In fact, resonant absorption occurs when a protein molecule absorbs EM radiation of a frequency that matches the natural  frequency of the molecule~\cite{prohofsky2004rf}. Resonant absorption is often  used in spectroscopy to identify proteins, as each protein has a unique set of natural frequencies. 

Subsequent to resonant absorption, proteins convert the absorbed EM energy into heat, which leads to a temperature increase within the protein and the surrounding medium.  Experiments of peptide chains in organic solvent indicate that a significant amount of dissipation can occur through the molecule-solvent interface~\cite{lervik2010heat}. In fact, the thermal conductivity of proteins is low compared to the thermal conductivity of water. Hence, proteins should be able to sustain
large thermal gradients across their structure, which may have implications on biological processes~\cite{lervik2010heat}. It is to be noted that the relationship between resonant absorption and thermal emission in proteins is complex and depends on a variety of factors, including the protein structure, the frequency of the incident EM wave, and the temperature gradient at the protein-water interface~\cite{prohofsky2004rf}. 

\begin{table*}[t]
	\footnotesize
	\centering
 	\caption{Comparison between Protein Modulation Techniques}
    \label{table:comparison}   
	\begin{tabular}{|p{0.15\textwidth}|p{0.4\textwidth}|p{0.2\textwidth}|p{0.07\textwidth}|}
		\hline
		\textit{\textbf{Technique}} & \textit{\textbf{Description}} & \textit{\textbf{Limitation}} & \textit{\textbf{Ref.}} \\
		\hline
		Optogenetics & 
        \begin{footnotesize}
- Optogenetics is a method that uses light to modulate molecular events in a targeted manner in living cells. 

- It relies on the use of genetically-encoded proteins that change conformation in the presence of light to alter cell behavior.
        \end{footnotesize} & 
        \begin{footnotesize}
- Requires genetic modification.

- Limited penetration depth.

- Limited accessibility.

- Limited temporal resolution.
        \end{footnotesize} & 
        \begin{footnotesize} 
\cite{fenno2011development} 
        \end{footnotesize} \\
        \hline
        Chemogenetics & 
        \begin{footnotesize}
- Chemogenetics uses small molecules called designer receptors, which are engineered to only respond to a specific drug.

- These molecules are used to selectively activate or inhibit specific groups of neurons, allowing researchers to target specific neural circuits in the brain.
        \end{footnotesize} & 
        \begin{footnotesize}
- Limited specificity.

- Non-reversible effects.

- Systemic effects.

- Variable responses.
        \end{footnotesize} & 
        \begin{footnotesize} 
\cite{gomez2017chemogenetics} 
        \end{footnotesize} \\
        \hline    
        Optogenomic & 
        \begin{footnotesize}
- Optogenomic interfaces are light-mediated nano-bio interfaces that allow the control of the genome, i.e., genes and their interaction in the cell nucleus.
        \end{footnotesize} & 
        \begin{footnotesize}
- Limited to in-vitro studies.

- Still at fundamental stage.
        \end{footnotesize} & 
        \begin{footnotesize} 
\cite{jornet2019optogenomic} 
        \end{footnotesize} \\
        \hline 
	\end{tabular}
\end{table*}

THz induced temperature changes may provide an alternative mechanism for modulating protein activities with several advantages over the already existing techniques. Table~\ref{table:comparison} provides a comparison between the various methods available in the literature highlighting their limitations and the advantages of using THz radiation as a protein modulation mechanism. In fact, THz-EM technology enables non-contact, non-invasive interrogation of biological tissues and neural circuits. Unlike optogenetics, which requires the introduction of light-sensitive proteins into cells through genetic manipulation~\cite{fenno2011development}, and chemogenetics, which involves the administration of exogenous ligands~\cite{gomez2017chemogenetics}, THz-EM technology does not require direct physical intervention, reducing potential damage and experimental artifacts.

In addition, THz-EM technology operates over a wide frequency range, spanning from a few hundred GHz to several THz. This broadband capability enables the detection of molecular vibrations, rotational transitions, and other spectral features associated with biological molecules, offering rich information about the composition, structure, and function of biological entities. Further, THz-EM technology can penetrate biological tissues deeply, allowing imaging and manipulation of internal structures without the need for invasive procedures. On the other hand, optogenetics is limited by the depth of light penetration, making it more suitable for superficial tissues or organisms with transparent or optically accessible body parts. In the case of chemogenetics, the penetration depth depends on the route of ligand administration, with limitations in reaching deep tissues compared to THz-EM technology.

Changes in temperature have the ability to modify the functioning of neurons by either altering membrane capacitance or stimulating temperature-sensitive ion channels~\cite{kim2022thermal}. Recall that in our system, the THz signal acts as an external stimulus that triggers the protein vibrational modes. Due to the absorption of the EM radiation, heat is the local stimulus that affects the temperature-sensitive ion channels.

\subsection{Mathematical Model}

Applying an external EM field to biomolecules causes a redistribution of charges within the molecule in relation to the field lines~\cite{romanenko2017interaction}. This creates a frequency-dependent permittivity response in the protein. Recall that permittivity is a measure of the electric polarizability of a dielectric. By analyzing the protein's response, changes in the protein structure at the THz frequency range can be explained by either relaxational or resonant processes. The relaxational response occurs in the amino acid side chains, while the resonant response comes from the coordinated movements of the protein structure~\cite{knab2006hydration}. When many parts of the system are coupled and oscillate simultaneously,  correlated motion arises. This implies that the nanoantenna can be adjusted to the protein's vibrational frequency to trigger targeted functional resonant interactions. 

The protein permittivity formulated as, $\varepsilon_p(\omega)=\varepsilon_p{'}(\omega)-j \varepsilon_p{''}(\omega)$, has its real part given by~\cite{knab2006hydration}
~\begin{equation}
\varepsilon_p{'}(\omega)=\varepsilon_{\infty}+\frac{ \varepsilon_{o}-\varepsilon_{\infty}}{1+(\omega \tau)^{2}}+\frac{(\varepsilon_{1}-\varepsilon_{\infty})\left[1-\left( \frac{\omega}{\omega_{o}} \right)^{2}\right]}{\left[ 1-\left( \frac{\omega}{\omega_{o}} \right)^{2} \right]^{2}+ (\omega\gamma)^{2}},
\end{equation}
while the imaginary part yields
\begin{equation}
\varepsilon_p{''}(\omega)=\frac{(\varepsilon_{o}-\varepsilon_\infty)\omega \tau}{1+(\omega \tau)^{2}}+\frac{(\varepsilon_{1}-\varepsilon_\infty)\omega \gamma}{\left[ 1-\left( \frac{\omega}{\omega_{o}} \right)^{2} \right]^{2}+ (\omega \gamma)^{2}}.
\end{equation}
Here, $\tau$ is the protein relaxation time, $\omega_o$ is  the natural   frequency of the protein, and $\gamma$ is the damping constant. $\varepsilon_{\infty}$ is the permittivity at the high-frequency limit, $\varepsilon_{o}$ is the relative permittivity at low frequencies (static region) and $\varepsilon_{1}$ refers to an intermediate permittivity value.  The damping constant $\gamma$ governs the magnitude of the resonances. The presence of damping limits the speed or rate at which proteins can fold, affecting their overall dynamics and behavior. 

In addition, the permittivity is related to  the complex refractive index $n(\omega)$  as follows
\begin{equation}
\varepsilon_{p}(\omega)=n^{2}(\omega)=(n'(\omega)+jn''(\omega))^2.
\label{eq:relative_perm}
\end{equation}
In~\eqref{eq:relative_perm}, $n'(\omega)$ and $n''(\omega)$ represent the refractive index real and imaginary parts, respectively. The  real part of the refractive index, which expresses the phase velocity, is  given as
\begin{equation}
n'(\omega)=\sqrt{\frac{1}{2}\left( |\varepsilon_{p}(\omega)|+\varepsilon_{p}{'}(\omega) \right)}.
\label{eq:realn}
\end{equation}
The imaginary part of the refractive index, which determines the attenuation of the EM wave, is expressed by
\begin{equation}
n''(\omega)=\sqrt{\frac{1}{2}\left( |\varepsilon_{p}(\omega)|-\varepsilon_{p}{'}(\omega) \right)},
\end{equation}
where $|\varepsilon_{p}(\omega)|$ is the complex modulus given as $\sqrt{\varepsilon_{p}^{'2}+\varepsilon_{p}^{''2}}$. 
 
Transmitted EM waves shift molecules in the medium to higher energy states resulting in molecular absorption~\cite{jornet2011channel}. Such a feature can be described using the absorption coefficient, $C_{abs}$. The molecular absorption coefficient of the particles can be  calculated using
\begin{equation}
C_{abs}(\omega)=\frac{4\pi n''(\omega)}{\lambda_{g}},
\label{eq:molecular absorption}
\end{equation}
where $\lambda_{g}$, the effective wavelength, is $\lambda /n'(\omega)$.

\section{Heat Transfer Model}
\label{Sec:Sec3}
Understanding thermal transport from nanoscale heat sources is important for a fundamental description of energy flow in materials, as well as for many technological applications including protein-mediated thermal therapies. Nevertheless, thermal transport at the nanoscale is fundamentally different from that at the macroscale and is determined by the energy dispersion in the material, the length scales of the heat sources, and the distance over which heat is transported~\cite{hoogeboom2015new}.

There are several factors that significantly affect the energy balance in heat transfer procedures. To begin with, the absorption coefficient of the biological structure (protein, cell, or tissue) at THz frequencies determines how much of the incident THz energy is absorbed as it propagates through it. Structures with higher absorption coefficients absorb more THz energy, leading to increased heating. The absorption coefficient depends on frequency, and this relationship is exhibited via the frequency-dependent refractive index.

Moreover, the power density, which refers to the amount of power per unit area, is a crucial parameter as it quantifies the concentration of EM energy within the structure being irradiated. Controlling the power density is important to ensure that the energy delivered remains within safe limits to prevent thermal damage or other adverse effects. In addition, the duration of exposure to THz radiation affects the amount of energy absorbed by the protein and the resulting temperature rise. Prolonged exposure may lead to cumulative heating effects, potentially causing thermal damage to the targeted region. Finally, the thermal conductivity of the targeted structure determines its ability to conduct heat away from the site of absorption. Biological structures with higher thermal conductivity may distribute heat more effectively, potentially mitigating localized heating effects.

While proteins and nanoparticles are fundamentally different in their composition, they do reveal similarities in their overall size, charge, and shape. In fact, the exterior surfaces of nanoparticles can be coated with organic functional groups which suggest that they could function as protein mimics~\cite{kotov2010inorganic}. Both theoretical calculations and experimental measurements have been conducted to better understand the heating process of nanoparticles and its effects on the surrounding medium. 

 Several studies on  both single and multiple nanoparticle heating under both continuous and pulsed illumination have been presented~\cite{govorov2007generating,baffou2011femtosecond,baffou2013photoinduced}. Hence, given the similarity between protein molecules and nanoparticles along with the vast literature on heat transfer models used for nanoparticles, we adopt the Goldenberg and Tranter heat transfer model~\cite{goldenberg1952heat,kang2021computational,xie2022single}. We specifically deploy this model as it provides the time-dependent solution to the heat conduction problem involving a heated sphere situated within a medium with distinct thermal characteristics. In dynamic systems, temperature variations occur continuously over time. Thereby, transient heat solutions allows us to understand the time evolution of temperature profiles.

In this paper, we consider protein molecules immersed in cytoplasm (modeled as water) and distributed on a cell surface. The Tx is a nano-laser impinging on protein molecules, which can be treated as heat sources.   To model our system, we consider one-dimensional transient heat transfer in spherical coordinates, where the governing equations are given by~\cite{kang2021computational}

\begin{equation}
\frac{1}{\alpha_p}\frac{\partial T_p(r,t)}{\partial t}=\frac{1}{r^2}\frac{\partial }{ \partial r}\left( r^2 \frac{\partial T_p}{\partial r} \right)+\frac{Q_v}{k_p}
\label{eq:firstpart}
\end{equation}
\begin{equation}
\frac{1}{\alpha_m}\frac{\partial T_m(r,t)}{\partial t}=\frac{1}{r^2}\frac{\partial }{ \partial r}\left( r^2 \frac{\partial T_m}{\partial r} \right).
\label{eq:secondpart}
\end{equation}
Here, $T$ is the temperature, $k$ is the thermal conductivity, $\alpha$ is the thermal diffusivity, $r$ is position and $t$ is time. The subscripts $m$  and $p$ indicate the medium and the protein, respectively.  

The temperature is uniform within the protein and can be considered as uniform volumetric heating, $Q_v$,  which is expressed using
\begin{equation}
Q_v=\frac{C_{abs}I}{v_p },
\end{equation}
where $I$ is the nano-laser power density, $v_p$ is the volume of the protein and $C_{abs}$ is the molecular absorption coefficient given in~\eqref{eq:molecular absorption}. In addition, the boundary  and initial conditions are given as~\cite{kang2021computational}
\begin{equation}
\begin{split}
\frac{\partial T}{\partial r}\big|_{r=0}&=0,\\
T\big|_{r=\infty}&=T_{\mathbf{ref}}=37\ \mathrm{^\circ C},\\
T\big|_{t=0}&=T_{\mathbf{ref}}=37\ \mathrm{^\circ C}.
\end{split}
\end{equation}

We also neglect the thermal resistance at the interface between the protein and the medium. The thermal resistance can be neglected since the conducting path for heat transfer is significantly longer compared to the inter-facial region. As such, the thermal resistance introduced by the interface becomes relatively small. This will result in the same temperature for the protein and the water immediately next to the protein surface. The continuous inter-facial conditions are described using~\cite{kang2021computational}
\begin{equation}
\begin{split}
T_p\big|_{r=R_p^-}&=T_m\big|_{r=R_p^+}\\
k_p\frac{\partial T_p}{\partial r}\big|_{r=R_p^-}&=k_m\frac{\partial T_m}{\partial r}\big|_{r=R_p^+},
\end{split}
\end{equation}
where $R_p$ is the protein  radius.

\begin{table*}
\centering
\begin{minipage}{0.75\textwidth}
\begin{align}
T_p(r,t)|_{t<\tau}= T_{ref}+\frac{R^2_pC_{abs}I}{k_pv_p}\Biggl\{\frac{k_p}{3k_m}+\frac{1}{6}\left(1-\frac{r^2}{R^2_p}\right)-\frac{2R_pb}{r\pi}  \int^{\infty}_{0}\frac{\exp(\frac{-y^2t}{\gamma_1})}{y^2} \frac{(\sin{y}-y\cos{y})\sin{(\frac{ry}{R_p})}}{\bigr[(c\sin y-y\cos y)^2+b^2y^2\sin^2 y\bigr]}dy\Biggl\}.
\label{eq:long1}
\end{align}
\end{minipage}
\end{table*}

\begin{table*}
\centering
\begin{minipage}{0.75\textwidth}
\begin{align}
T_m(r,t)|_{t<\tau}= T_{ref}+\frac{R^3_pC_{abs}I}{rk_pv_p}\Biggl\{\frac{k_p}{3k_m}-\frac{2}{\pi} \int^{\infty}_{0}\frac{\exp(\frac{-y^2t}{\gamma_1})}{y^3}\frac{(\sin{y}-y\cos{y})\bigr[(by\sin y\cos\sigma y)-(c\sin{y}-y\cos{y})\sin\sigma y\bigr]}{\bigr[(c\sin y-y\cos y)^2+b^2y^2\sin^2 y\bigr]}dy\Biggl\}.
\label{eq:long2}
\end{align}
\medskip
\end{minipage}
\end{table*}

\begin{table*}
\centering
\begin{minipage}{0.75\textwidth}
\begin{align}
b=\frac{k_m}{k_p}\sqrt{\frac{\alpha_p}{\alpha_m}}, c=1-\frac{k_m}{k_p}, \gamma_1=\frac{R^2_p}{\alpha_p}, \sigma=\Bigg(\frac{r}{R_p}-1\Bigg)\sqrt{\frac{\alpha_p}{\alpha_m}}.
\label{eq:long3}
\nonumber
\end{align}
\medskip
\end{minipage}
\end{table*}
The complete form of the analytical solutions of~\eqref{eq:firstpart} and~\eqref{eq:secondpart} are given in~\eqref{eq:long1} and~\eqref{eq:long2}~\cite{kang2021computational}. Laplace transformation methods yield the formal solution, which
provides the time-dependent temperature for an unlimited long pulse. Since the differential equations for heat diffusion are linear, a solution for a rectangular laser pulse width $\tau$ can be constructed by subtracting the solutions of $T(t, r)$, which are separated in time, as follows~\cite{liu2009mechanisms} 
\begin{equation}
\Delta T= T(r,t)-T(r,t-\tau).
\end{equation}

For a distribution of proteins, the temperature profile can be found using the superposition of the analytical solution for single protein heating as follows~\cite{xie2022single}
\begin{equation}
T_{MP}(r,t)=\sum^{N}_{i=1}T^i_p(r^i,t),
\label{eq:multiple_protein}
\end{equation}
where $T_{MP}$ is the temperature profile for multiple particle heating and  $T^i_p$ is the analytical solution for the $i^{th}$ single-particle heating given in~\eqref{eq:long1}. 

It is important to recognize that proteins operate within broader networks and complexes. Consequently, when aiming to induce a desired temperature alteration in our system, we take into account a population of proteins. This distinction is crucial when comparing thermal studies on proteins and nanoparticles, as the latter may not necessarily function as constituents of larger networks.

Using~\eqref{eq:long1}, we can also find the temperature change experienced by a protein particle in steady-state as follows
\begin{equation}
\Delta T_{ss}=\frac{C_{abs}I}{4 \pi k_m R_p}.
\end{equation}
Finally, when a thermal perturbation is applied at some point in a medium (e.g., an instantaneous change in a surface temperature), it generally takes on the order of $\tilde t$ for the perturbation to appear at a distance from the particle. Thus, the time scale related to the thermal process, $\tilde t$, is given by~\cite{minkowycz2012nanoparticle}
\begin{equation}
\tilde t= \frac{d^2_{pp}}{\alpha_p},
\end{equation}
where $d_{pp}$ is the average particle-particle separation.
 
It is worth mentioning that the assumption of uniform heat distribution becomes less valid as the size of the protein cluster increases relative to the size of the heating source or the system as a whole. In cases where the protein cluster is significantly larger than the heating source, the heat may not be distributed uniformly throughout the protein cluster. Instead, localized heating effects may dominate, leading to non-uniform temperature distributions. The violation of the assumption of uniform heat distribution is more likely to occur when the size of the protein cluster approaches or exceeds the characteristic length scale associated with the heating process. In general, there is no strict cutoff size at which the assumption of uniform heat distribution is violated. Instead, the validity of this assumption depends on various factors, including the relative sizes of the heating source, the protein or protein cluster, and the surrounding environment, as well as the efficiency of heat transfer mechanisms within the system.
 
 Arbitrary protein shapes  can affect the assumption of uniform heat distribution, particularly when considering non-symmetric or complex geometries. In such cases, heat conduction may not occur uniformly in all directions, leading to non-uniform temperature distributions within the system. To account for these effects and modify the governing equations accordingly, one should incorporate anisotropic heat conduction terms into the heat transfer equations to account for directional variations in heat propagation. In addition, complex boundary conditions should be considered to reflect the irregular shape of the system.

\section{Ion Channels Gated by Heat}
\label{Sec:Sec4}

Temperature-sensitive ion channels are a class of ion channels that respond to changes in temperature by altering their ion conductance and influencing cellular excitability. These channels play a crucial role in temperature-dependent physiological processes. Below are some examples of temperature-sensitive ion channels.

\subsection{TRP Channels}

A unique group of temperature-sensitive ion channels, named Transient Receptor Potential (TRP) channels (also referred to as thermoTRP), has been recognized as primary bio-thermometers in a range of species. They are found in various tissues and organs, including sensory neurons, skin cells, and internal organs, where they help to detect and transmit temperature signals to the brain. Direct gating by temperature changes implies that the channel protein is able to absorb heat and convert it to a conformational change efficiently. Each thermoTRP channel has its own specific range of temperatures at which it is most effectively activated. These distinct temperature ranges have led to the concept of an activation threshold for each channel~\cite{emir2017neurobiology}. In fact, the Transient Receptor Potential Vanilloid (TRPV) channels are a subgroup of the larger TRP channel family. Within this subgroup, the TRPV1 is sensitive to noxious heat, the TRPV2 responds to higher temperatures, the TRPV3 is activated by moderate warmth, and the TRPV4 is involved in sensing warm and tepid temperatures.

\subsection{K2P Channels}
Some members of the two-pore domain potassium (K2P) channel family, such as TREK-1 and TRAAK, are considered thermosensitive. In fact, TREK-1 is a potassium ion channel that can be modulated by temperature. Recall that potassium ion channels are highly specialized proteins that exhibit specific structural and functional properties. The gating mechanism of TREK-1 involves sensitivity to changes in temperature, leading to alterations in its conformation and activity. When the temperature increases, TREK-1 channels tend to open, allowing the flow of potassium ions across the cell membrane. The precise molecular details of how temperature affects the conformation and gating of TREK-1 are still being investigated, yet it is believed that changes in temperature directly influence the lipid bilayer surrounding the channel, leading to alterations in its structure~\cite{lamas2019ion}.

Similarly, TRAAK channels are potassium channels that can be influenced by changes in temperature, specifically by heat. 
Much like the other members of the family, TRAAK currents showed a strong open-channel outward rectification when recorded in whole-cell configuration, and these currents increase strongly when the temperature rises. As such, heat can modulate the gating of TRAAK channels, affecting their open and closed states. The exact temperature range at which TRAAK channels are most sensitive and their precise gating behavior in response to heat can vary among different cell types and experimental conditions. Nevertheless, the heat sensitivity of TRAAK channels allows them to serve as key regulators of cellular excitability and contribute to thermosensation processes in various tissues and organisms~\cite{lamas2019ion}.

\subsection{HCN Channels}

Hyperpolarization-activated Cyclic Nucleotide-gated (HCN) channels are a class of ion channels found in various neurons, including those in the brain and the heart. HCN channels are unique in that they are primarily activated by hyperpolarization, but they also exhibit sensitivity to temperature, particularly heat. In fact, heat can enhance the opening of HCN channels, leading to an influx of positively charged ions, such as sodium (Na+) and potassium (K+), into the cell. This inward current can depolarize the cell membrane and influence neuronal excitability~\cite{ramentol2020gating}.

In addition, HCN channels play a critical role in pacemaker activity, which is responsible for generating rhythmic electrical signals in certain neurons, including those in the heart (cardiac pacemaker cells) and the brain (pacemaker neurons). Heat activation of HCN channels can contribute to the regulation of the pacemaker activity by adjusting the rate and frequency of action potential generation~\cite{ramentol2020gating}.

Finally, the  dysfunction of HCN channels can have pathological consequences. Altered HCN channel activity due to genetic mutations or changes in temperature sensitivity may contribute to disorders such as epilepsy, cardiac arrhythmia, and neuropathic pain~\cite{ramentol2020gating}.

\subsection{Opening Probability of Thermally-Gated Channels}
For thermally-gated ion channels, the channel opening reaction can be obtained by considering a two-state model. The equilibrium between the open ($O$) and closed ($C$) state of the channel can be described by the following constant~\cite{voets2012quantifying}
\begin{equation}
K=\frac{O}{C}=\exp\Bigg({\frac{-\Delta H+T\Delta S+zFV}{RT}}\Bigg),
\end{equation}
where $\Delta H$ is the difference in enthalpy between the ($O$) and ($C$) states (J mol$^{-1}$) and $\Delta S$ is the difference in entropy  between the ($O$) and ($C$) states (J mol$^{-1}$ K$^{-1}$), respectively. $z$ is the gating charge (which equals 0 in
the case of a voltage-independent channel), $V$ is the transmembrane voltage, $F$ is the
Faraday constant and $R$ is the universal gas constant. 

The channel’s open probability is given by~\cite{voets2012quantifying}
\begin{equation}
P_{open}=\frac{O}{O+C}=\frac{1}{1+\frac{1}{K}}=\frac{1}{1+\exp\bigg({\frac{\Delta H-T\Delta S-zFV}{RT}}\bigg)}.
\label{eq:main_eq}
\end{equation}
We can notice from~\eqref{eq:main_eq} that the open probability changes with temperature according to~\cite{voets2012quantifying}
\begin{equation}
\begin{split}
\frac{dP_{open}}{dT}&= \Bigg(\frac{1}{1+\exp\bigg({\frac{\Delta H-T\Delta S-zFV}{RT}}\bigg)}\Bigg)^2 \times \\  &\exp\bigg({\frac{\Delta H-T\Delta S-zFV}{RT}}\bigg) \times \frac{(\Delta H-zFV)}{RT^2}.
\end{split}
\end{equation}
It follows that a
channel is heat-activated (i.e. $\frac{dP_{open}}{dT} > 0$) when $(\Delta H-zFV) > 0$. Finally, in the case of thermally-gated channels that are both temperature and voltage-dependent, as is the case for the TRPV1 channel,~\eqref{eq:main_eq} can be rewritten as
\begin{equation}
P_{open}=\frac{1}{1+\exp({\frac{zF}{RT}}(V_{1/2}-V))},
\label{eq:main_eq1}
\end{equation}
where $V_{1/2}$ represents the voltage for half-maximal activation. $V_{1/2}$ changes linearly with temperature according to
\begin{equation}
V_{1/2}=\frac{1}{zF}(\Delta H-T\Delta S).
\end{equation}
We note that in these equations, the temperature $T$ is given by~\eqref{eq:long1} and~\eqref{eq:long2}. 
\section{Results}
\label{Sec:Sec5}
The theoretical model presented in this work is simulated using both MATLAB and COMSOL Multiphysics\textsuperscript{\textregistered}, where the simulation parameters are given in Table~\ref{table:sim_parameters}.  In our model, we assume that the protein particles are distributed on the surface of a cell and embedded in an aqueous solution mimicking the cell cytoplasm.  The surface of the cell has an area of 10~$\mu$m~$\times$~10~$\mu$m. The cell is assumed to be fixed since the main focus of the paper is on the interaction between the THz source and the protein patch on the cell surface. In another work we presented earlier~\cite{elayan2023terahertz}, we have studied the impact of cell motion on protein folding.

We perform the simulations on the protein lysozyme, which is characterized in Table~\ref{table:protein_perm_values}. We assumed only one type of proteins (only lysozyme), as the aim of the work is to demonstrate how the interaction of a THz laser beam with a group of lysozyme proteins can generate controlled and precise temperature changes based on the values of the signal power, pulse duration, and inter-particle distance. In addition, we assume that each protein is a standing sphere with a radius of 5~nm. In the context of this work, modeling the protein as a sphere is useful since spherical models provide rough estimates of properties such as volume, surface area, and diffusion coefficients, which are important in various biochemical and biophysical studies. For the cytoplasm, we use the parameters of water at THz as provided in~\cite{reid2013terahertz}. 

 We note that for THz induced temperature to impact protein molecules, the accumulation of heat fluxes produced by individual particles is required to create a measurable, cumulative temperature effect. 
Larger clusters with a higher density of proteins are likely to generate more heat and exhibit more pronounced effects on the temperature distribution compared to smaller clusters (Collective heating vs. Localized heating).

\begin{table}[t]
	\footnotesize
	\centering
 	\caption{Simulation Parameters}
 \label{table:sim_parameters}   
	\begin{tabular}
		{|p{0.2\textwidth}|p{0.15\textwidth}|p{0.05\textwidth}|}
		\hline
		\hline
		\multicolumn{1}{|l|}{ \textit{\textbf{Parameter}}} & \multicolumn{1}{l|}{ \textit{\textbf{Value}}}  & \multicolumn{1}{l|}{ \textit{\textbf{Ref}}} \\
		\hline
		\hline
		Protein Thermal Conductivity & 0.1-0.2  W/m$\cdot$K 
        & \cite{lervik2010heat}
          \\
\hline
Water Thermal Conductivity & 0.6 W/m$\cdot$K 
        & \cite{lervik2010heat}
          \\
          \hline
Protein Specific Heat Capacity & 1483 J/kg$\cdot$K 
        & \cite{yang1979protein}
        \\
        \hline
Water Specific Heat Capacity & 4180 J/kg$\cdot$K 
        & -
        \\
         \hline
Protein Density & 1350 kg/m$^3$ 
& \cite{fischer2004average}
        \\
\hline     
Protein radius & 5 nm 
& \cite{erickson2009size}
        \\
        \hline
        Inter-particle distance & 10 nm
& \cite{erickson2009size}
\\
\hline
  Number of particles & 10$^4$
& -
\\
\hline
Pulse Duration & 1 millisecond (ms) &-
\\
\hline
	\end{tabular}
\end{table}

\begin{table}[t]
	\footnotesize
	\centering
 	\caption{Protein Permittivity Values~\cite{son2014terahertz}}
 \label{table:protein_perm_values}  
	\begin{tabular}
		{|p{0.1\textwidth}|p{0.03\textwidth}|p{0.03\textwidth}|p{0.03\textwidth}|p{0.03\textwidth}|p{0.05\textwidth}|p{0.05\textwidth}|}
		\hline
		\hline
		\multicolumn{1}{|l|}{ \textit{\textbf{Protein Type}}} & \multicolumn{1}{l|}{ \textit{\textbf{\begin{footnotesize}$\varepsilon_{\infty}$\end{footnotesize} }}} &\multicolumn{1}{l|}{ \textit{\textbf{\begin{footnotesize}$\varepsilon_{o}$\end{footnotesize} }}} & \multicolumn{1}{l|}{ \textit{\textbf{\begin{footnotesize}$\varepsilon_{1}$\end{footnotesize} }}}  & \multicolumn{1}{l|}{ \textit{\textbf{\begin{footnotesize}$\tau$(ps)\end{footnotesize} }}} &\multicolumn{1}{l|}{ \textit{\textbf{\begin{footnotesize}$\omega_o$(THz)\end{footnotesize} }}}&\multicolumn{1}{l|}{ \textit{\textbf{\begin{footnotesize}$\beta$(ps)\end{footnotesize} }}}\\
		\hline
		\hline
		Lysozyme & 1 
        & 216 & 0.73 & 400& 1.8 & 0.02
          \\
\hline
	\end{tabular}
\end{table}

\subsection{MATLAB Results}

\begin{figure}[h!]
\centering
\includegraphics[width=0.5\textwidth]{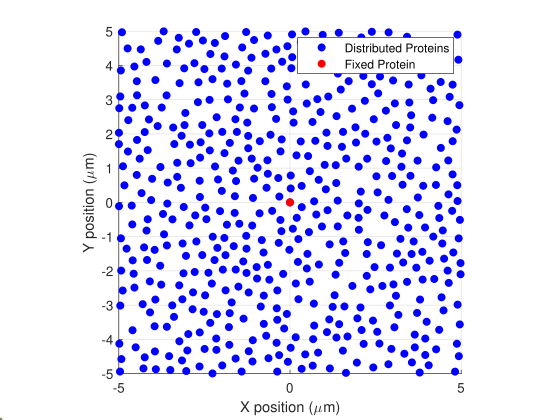}
\footnotesize
\caption{A visualization of the simulation setup.}
\label{fig:visual}
\end{figure}

For the MATLAB results, we numerically evaluate~\eqref{eq:long1},~\eqref{eq:long2} and~\eqref{eq:multiple_protein}. We generate random positions for the protein particles over the surface area of a single cell. We also allocate a protein particle in the center to be our point of reference. The distance from the center of the reference particle to the boundary of the area is 5~$\mu$m as shown in Fig.~\ref{fig:visual}. In our calculations, we utilize the dielectric response of lysozyme as a function of hydration, obtained from terahertz time-domain spectroscopic measurements. As such, all the necessary parameters, including the dispersion constant and the characteristic relaxation time for the system, have been extracted from~\cite{knab2006hydration}. Since the size of the scatterers at THz frequencies is much smaller than the wavelength of the propagating THz wave, the scattering effect is almost negligible compared to absorption, as demonstrated in the results presented in \cite{elayan2017terahertz}. This is why we only incorporated the THz absorption properties.

The nanowatt-microwatt power range represents a general expectation for efficient and practical operation in nanoscale systems. By operating at such power levels, nanonetworks can benefit from reduced energy consumption, enabling long-term autonomous operation, especially in scenarios where access to a continuous power source may be challenging or impractical. It also allows for the integration of nanoscale devices into energy-constrained environments, including wearable and implantable medical devices.
  
\begin{figure}[h!]
\centering
\includegraphics[width=0.45\textwidth]{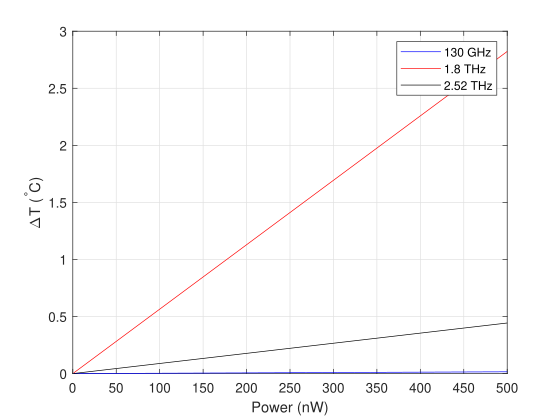}
\footnotesize
\caption{Temperature change versus nano-laser power for lysozyme protein particles.}
\label{fig:temp_vs_power}
\end{figure}

  Fig.~\ref{fig:temp_vs_power} demonstrates the temperature increase versus the nano-laser power for a pulse duration of 1~ms. A lower or higher temperature change can be achieved by controlling the nano-laser power level depending on the intended medical application. We note that thermal denaturation of proteins occurs when the temperature increase is $\geq4^\circ$C~\cite{lepock1993protein}. Hence, this value represents
a threshold that should not be exceeded for temperature increase. In fact, denaturation
affects the secondary and tertiary structure of proteins, including disruption of hydrogen
bonds, changes in the orientation of alpha-helices and beta-sheets, as well as changes in
the solvent-accessible surface area of the protein.
  
  It is important to mention that the frequencies compared in Fig.~\ref{fig:temp_vs_power} involve the 130~GHz, which is the most common frequency for wireless communication in the THz frequency band, the 1.8~THz, which is the resonant frequency of the protein lysozyme as well as the 2.52~THz which is a conventional frequency for THz imaging applications. 

\begin{figure}[h!]
\centering
\includegraphics[width=0.45\textwidth]{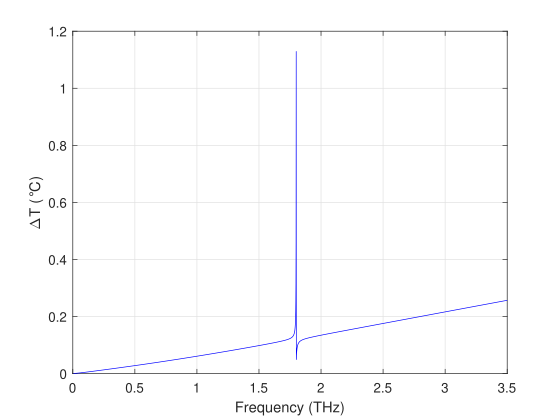}
\footnotesize
\caption{Temperature change versus frequency  for lysozyme protein particles.}
\label{fig:temp_vs_freq}
\end{figure}

Fig.~\ref{fig:temp_vs_freq} shows the temperature increase versus the nano-laser frequency at 200~nW for a pulse duration of 1~ms. It can be seen that at the natural frequency of the protein lysozyme, i.e. 1.8~THz, we experience the highest temperature increase.  The results reaffirm the importance of the resonant absorption phenomenon and its significance when studying the induced thermal effects of THz radiation. We clearly observe ``anomalous dispersion", which occurs when the refractive index decreases with an increase of frequency, due to the absorption of radiation. 

In scenarios where other protein types are present, the selection of the THz laser source depends on their frequencies. In general, resonance is a very powerful tool, so if the THz laser source is tuned to the vibrational frequency of the protein of interest, only the targeted protein group will be stimulated. In fact, in our previous work~\cite{9819908}, we showed that if the difference between the protein resonant frequencies available in the system is $\geq$ 0.2 THz, then selectivity is always maximum at resonance. This value has been attained after conducting several simulation trials and experimenting with different proteins of varying resonances. Nevertheless, if the difference is less than 0.2 THz, the overlap between the protein populations gets larger and more of the untargeted protein population can get impacted resulting in a higher number of false positives. In this case, we need to use the metric provided in~\cite{9819908}, as it directs us to the frequency at which the laser source should be tuned.

\begin{figure}[h!]
\centering
\includegraphics[width=0.45\textwidth]{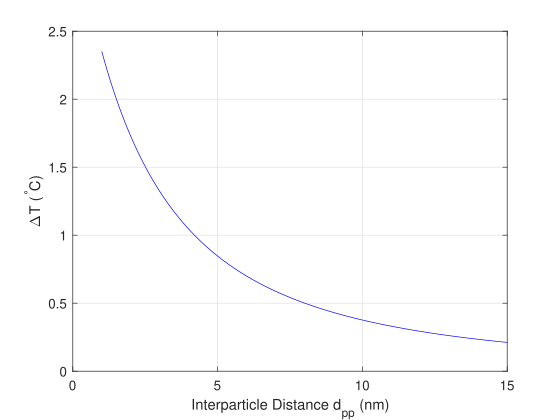}
\footnotesize
\caption{Temperature increase versus inter-particle distance.}
\label{fig:temp_vs_interdistance}
\end{figure}

Fig.~\ref{fig:temp_vs_interdistance} presents the temperature increase versus the inter-particle distance between the lysozyme protein particles for a Tx power of 200~nW and a pulse duration of 1~ms. It can be concluded that when protein particles are closer together, the interaction between their temperature fields is more significant because the thermal fields from neighboring particles overlap more quickly. 

\begin{figure}[h!]
\centering
\includegraphics[width=0.45\textwidth]{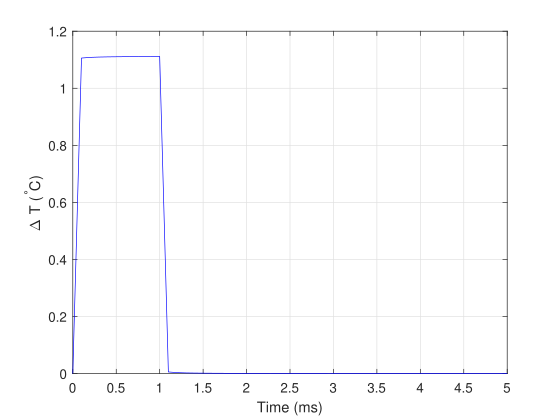}
\footnotesize
\caption{Temperature increase versus pulse duration.}
\label{fig:temp_vs_time}
\end{figure}

It is to be noted that rapid temperature sensing and channel response times are essential for modulating neuronal activity in response to temperature changes. Millisecond-scale timing ensures the appropriate integration of temperature signals into neuronal firing patterns and synaptic transmission. As such, Fig.~\ref{fig:temp_vs_time} illustrates the temperature change for lysozyme protein particles versus time, where a 1~ms pulse-width and a 200~nW Tx power are utilized.

\begin{table}[t]
	\footnotesize
	\centering
 	\caption{TRPV3 Parameters \cite{nadezhdin2021structural}}
 \label{table:TRPV_parameters}   
	\begin{tabular}
		{|p{0.15\textwidth}|p{0.15\textwidth}|}
		\hline
		\hline
		\multicolumn{1}{|l|}{ \textit{\textbf{Parameter}}} & \multicolumn{1}{l|}{ \textit{\textbf{Value}}}  \\
		\hline
		\hline
		$\Delta H$ & 91.2  kcal/mol 
          \\
\hline
$\Delta S$ & 287.5 cal/mol$\cdot$K 
          \\
          \hline
$z$ & 0.7~\cite{nilius2005gating}
        \\
\hline
	\end{tabular}
\end{table}

\begin{figure}[h!]
\centering
\includegraphics[width=0.45\textwidth]{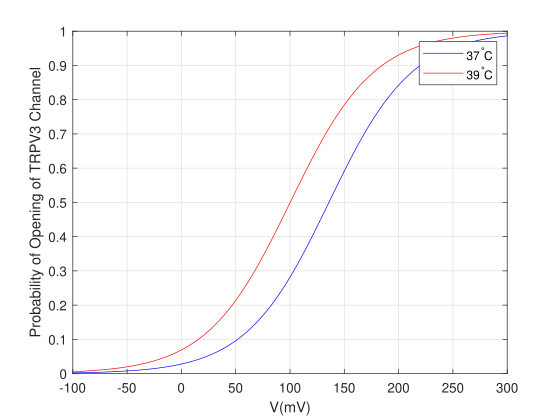}
\footnotesize
\caption{Steady-state opening probability of TRPV3 channel.}
\label{fig:thermal_channels}
\end{figure}

Finally, using the parameters in Table~\ref{table:TRPV_parameters}, we simulate the steady-state probability of a thermally activated channel, in this case the TRPV3 channel, as shown in Fig.~\ref{fig:thermal_channels}. Under resting conditions, i.e. at 37$^\circ$C, the midpoint of the activation curve occurs at  a  potential around 136~mV for a gating charge $z=0.7$~\cite{nadezhdin2021structural}. However, as the temperature increases, the activation voltage shifts towards the left, resulting in channel activation at 100~mV, which is closer to voltage levels that are more representative of normal cellular conditions.

 \subsection{COMSOL Model}

\begin{figure}[h!]
\centering
\includegraphics[width=0.4\textwidth]{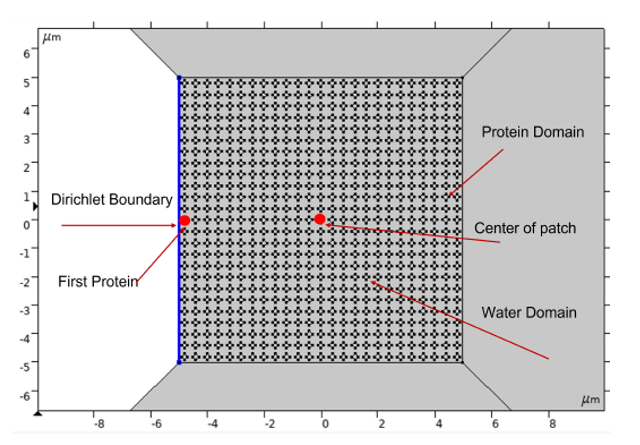}
\footnotesize
\caption{COMSOL model of THz signal incident on protein patch.}
\label{fig:comsol_model}
\end{figure}

 COMSOL Multiphysyics\textsuperscript{\textregistered} is used to model the thermal effect of a 1.8~THz-EM wave on a group of proteins.  To verify the analytical results, we built a 2D parametric model of a 10~$\mu$m~$\times$~10~$\mu$m patch of proteins in cytoplasm (modelled as water).  As such, the model is composed of two domains, one for water, and one for the protein molecules. Each domain has its own set of thermal as well as electrical parameters  as given in  Table~\ref{table:sim_parameters} and Table~\ref{table:protein_perm_values}. Similar to the MATLAB simulations, the source of radiation is a rectangular beam incident on the Dirichlet boundary as shown in Fig.~\ref{fig:comsol_model}. It is to be noted that the same surface area covered by the proteins in MATLAB is covered by the proteins in COMSOL. Basically, a refined mesh that accurately captures the system geometries has been implemented.

In this work, two different physics were considered, one for the wave propagation and the other for studying the temperature change through the ``bioheat transfer" physics module. The mechanism works as follows: the model feeds the power density distribution data to the heat transfer module for determining the temperature change in the defined domains.

\begin{figure}
 \centering
 \subfigure[]{%
  \includegraphics[height=4 cm, width=7cm]{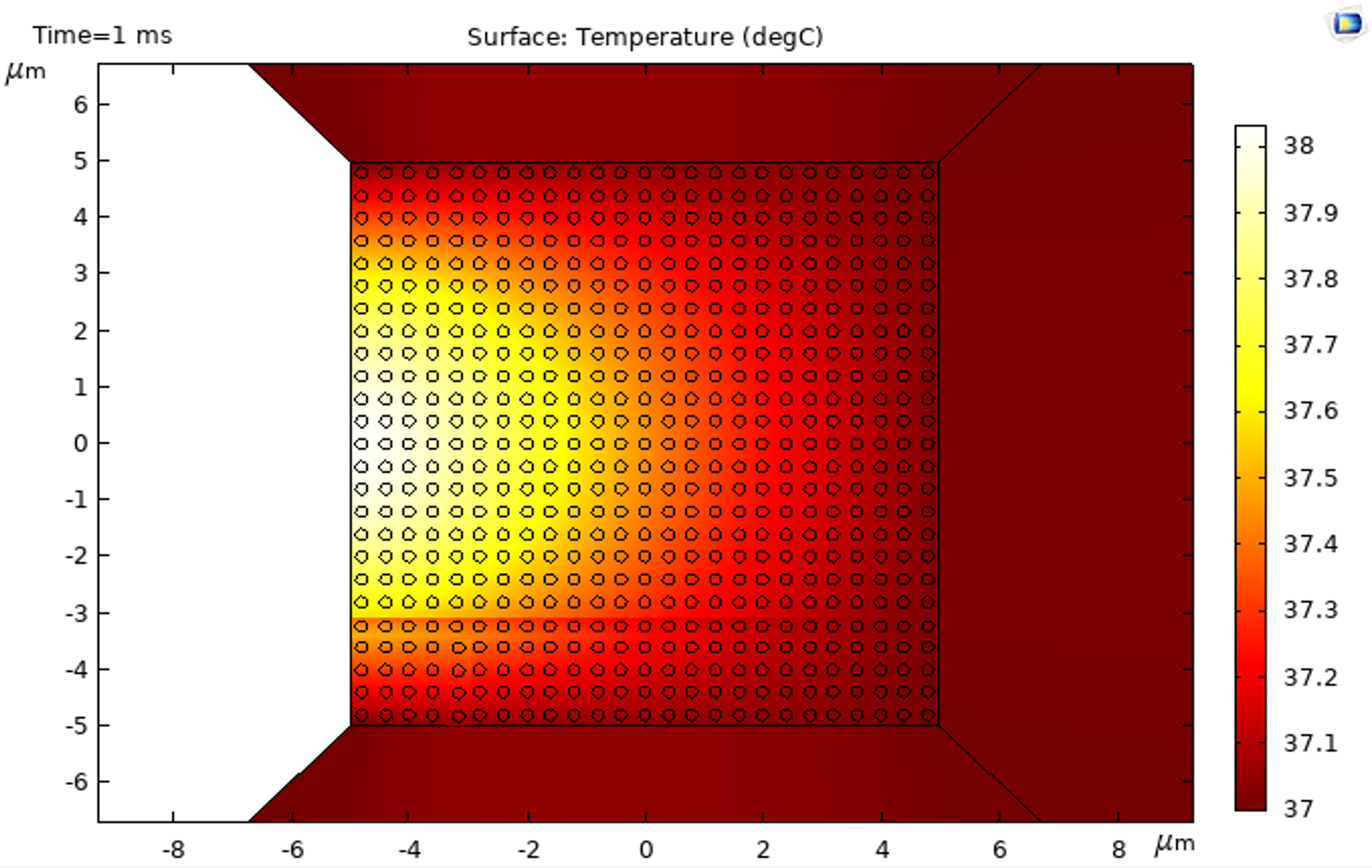}%
}

\subfigure[]{%
  \includegraphics[height=4 cm, width=7cm]{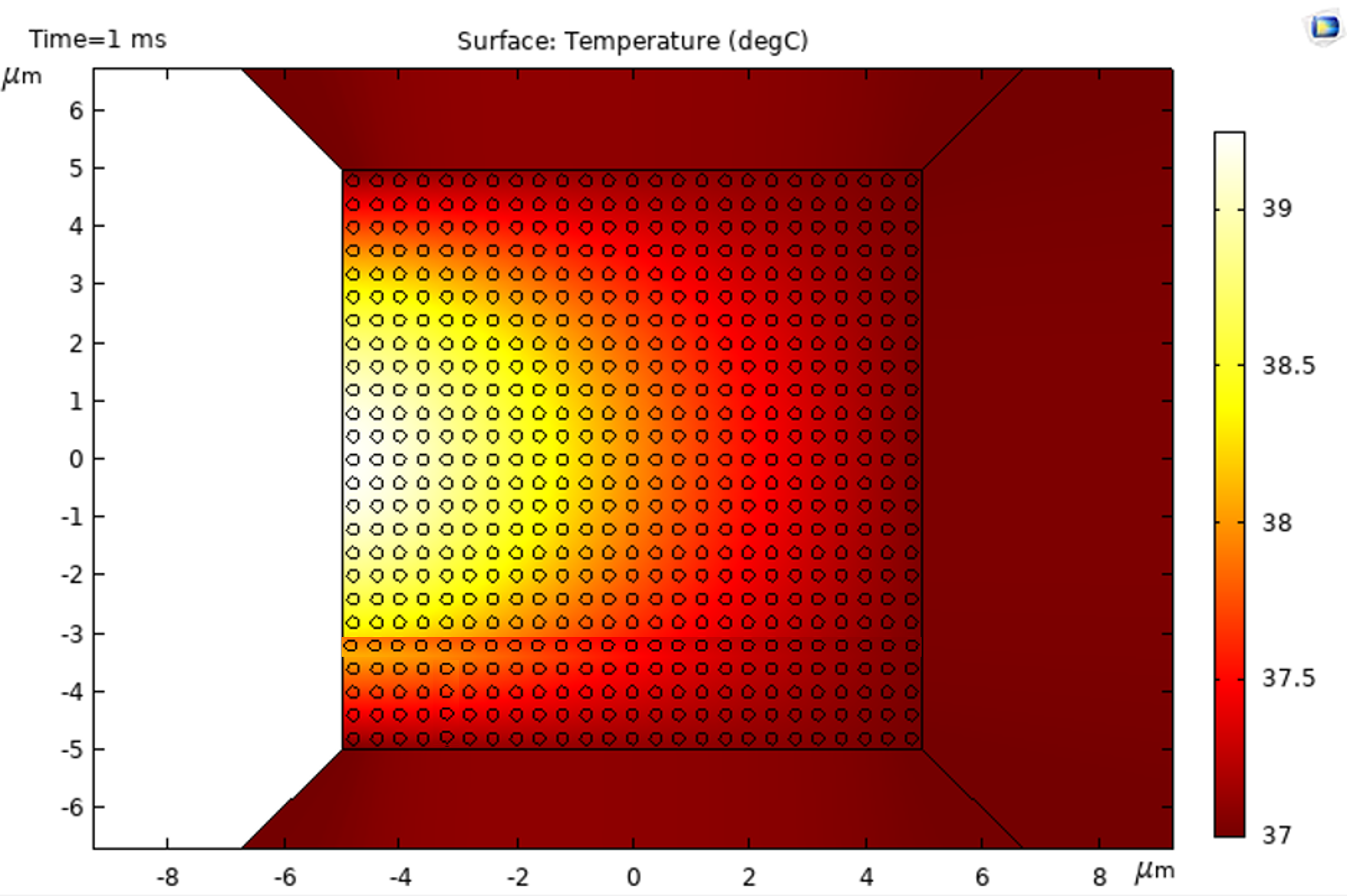}%
}

\caption{Protein patch targeted using (a) 200~nW laser beam for 1 millisecond. (b) 400~nW laser beam for 1 millisecond.}
\label{fig:comsol_model2}
\end{figure}

\begin{figure}
 \centering
 \subfigure[]{%
  \includegraphics[height=5 cm, width=8cm]{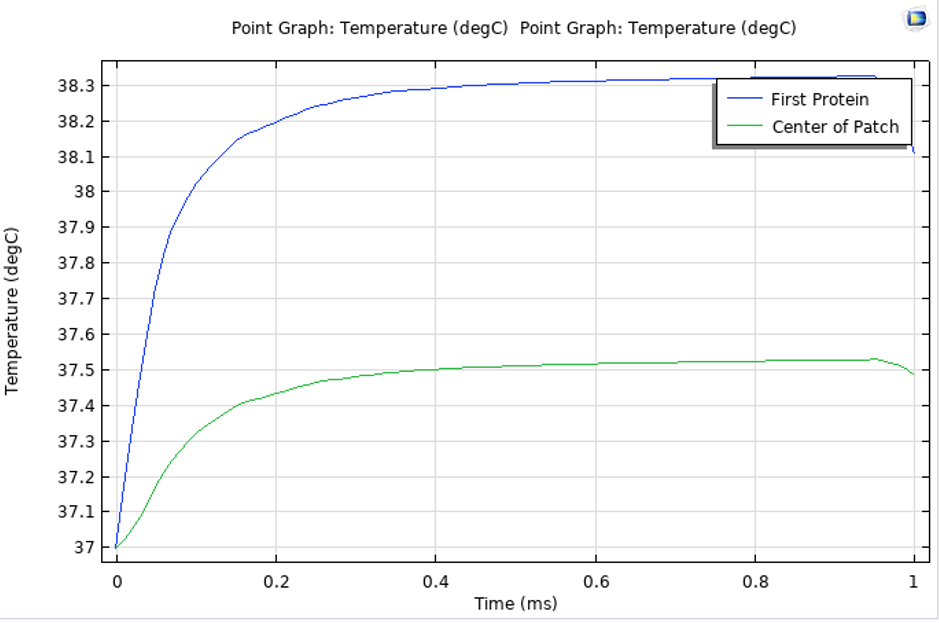}%
}

\subfigure[]{%
  \includegraphics[height=5 cm, width=8cm]{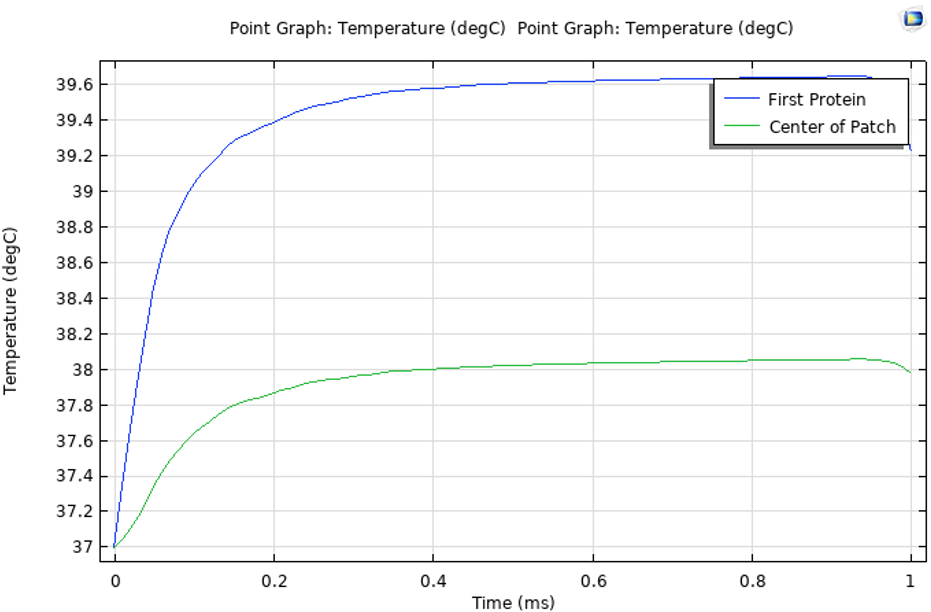}%
}

\caption{(a) Difference in temperature between the first protein and the center of the protein patch illuminated by (a) 200~nW power (b) 400~nW power.}
\label{fig:comsol_model3}
\end{figure}

Fig.~\ref{fig:comsol_model2} presents the interaction of a THz laser beam with a group of lysozyme proteins. We can notice from the figure that  the thermal energy is dispersed throughout the entire patch. Upon increasing the power from 200~nW in Fig.~\ref{fig:comsol_model2}a to 400~nW in Fig.~\ref{fig:comsol_model2}b, the temperature increase is 2$^\circ$C.  Comparing Fig.~\ref{fig:comsol_model2} with the simulation results in Fig.~\ref{fig:temp_vs_power},  we see that the two simulation environments are in agreement. 
 Hence, through THz induced thermal effects, we can generate controlled and precise temperature changes. 
In addition, Fig.~\ref{fig:comsol_model3} demonstrates the difference in temperature between the first protein impacted by the THz laser beam and the center of the patch for a 200~nW and a 400~nW power level, respectively. It can be concluded that the cell starts to heat up from the side
the laser radiates through it, and then the cell itself turns
into a heat source and dissipates the heat to the medium.

\subsection{Experimental Results}
Using the TeraNova testbed~\cite{sen2020teranova}, 
 an Arbitrary Waveform Generator (AWG) generates an intermediate frequency (IF) signal of 1 GHz. To up-convert the signal to 130 GHz, two separate frequency doublers (x4) and a mixer chain have been utilized. The signal is transmitted through a conical horn antenna with 10$^\circ$ half-power beam-width and 21 dBi gain. This device has two synchronized channels operating simultaneously with up to 32 GHz of analog bandwidth.

\begin{figure}[h!]
\centering
\includegraphics[width=0.51\textwidth]{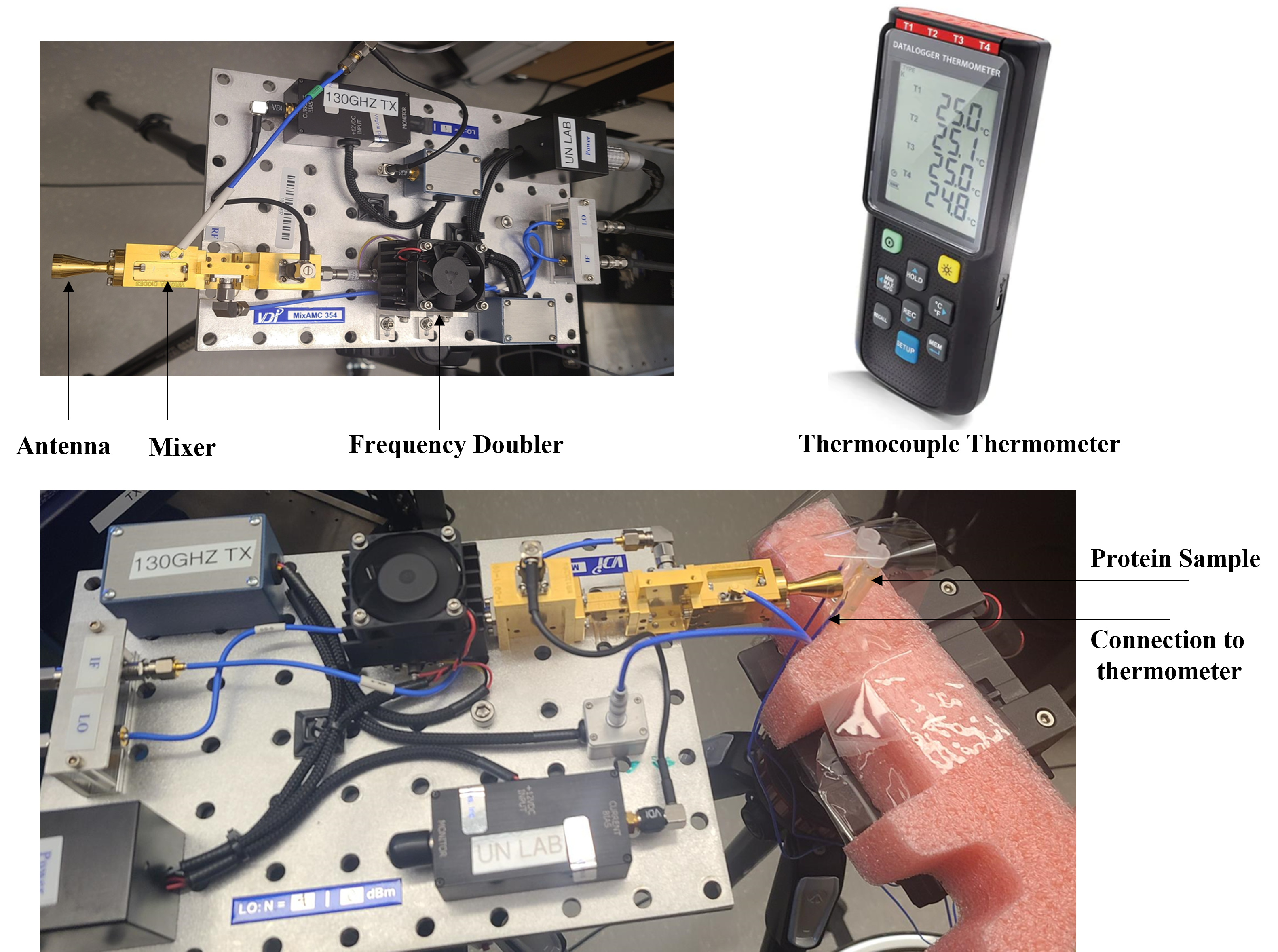}
\footnotesize
\caption{Components of the Experimental Setup}
\label{fig:setup2}
\end{figure}

\begin{figure}[h!]
\centering
\includegraphics[width=0.5\textwidth]{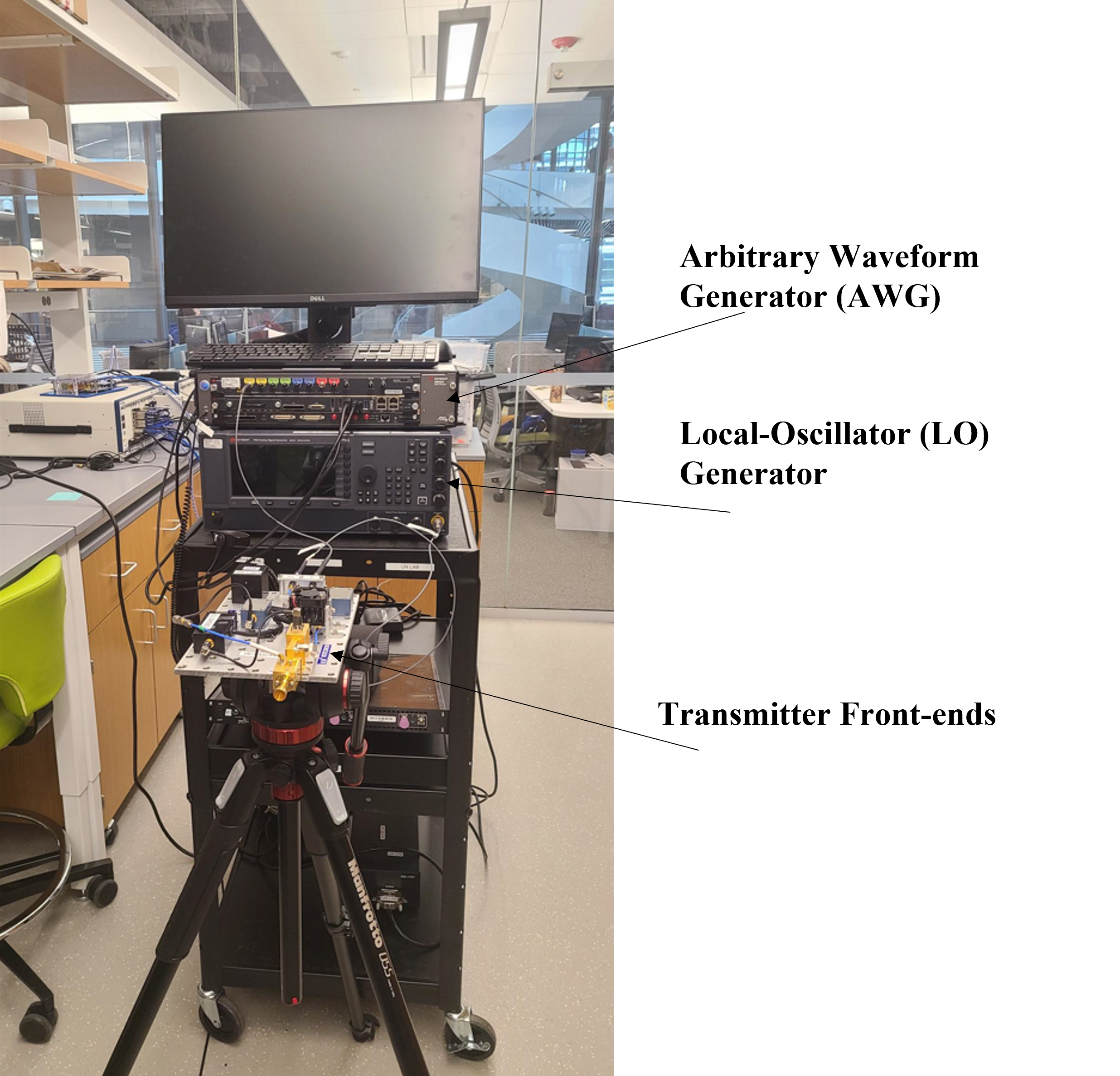}
\footnotesize
\caption{Complete Experimental Setup}
\label{fig:setup1}
\end{figure}

The components of the experimental setup are outlined in Fig.~\ref{fig:setup2}. A 130 GHz transmitter is used to radiate a signal in the sub-THz frequency band. At the receiver, a thermocouple-thermometer is used in order to measure the changes in temperature experienced by the protein sample. For this experiment, we have used a 2~mL of egg-white as it is an aqueous-rich medium mainly composed of proteins, such as ovalbumin, ovotransferrin, lysozyme and ovomucin. For comparison purposes, we also radiated a 2~mL sample of water with the 130 GHz Tx. As can be seen from Fig.~\ref{fig:setup2}, the antenna impinges directly on the protein sample to avoid losses. The distance between the transmitter and the radiated samples is 2~cm. The power transmitted is 45~$\mu$W.  A complete configuration of the experimental setup is provided in Fig.~\ref{fig:setup1}.

\begin{figure}[h!]
\centering
\includegraphics[width=0.45\textwidth]{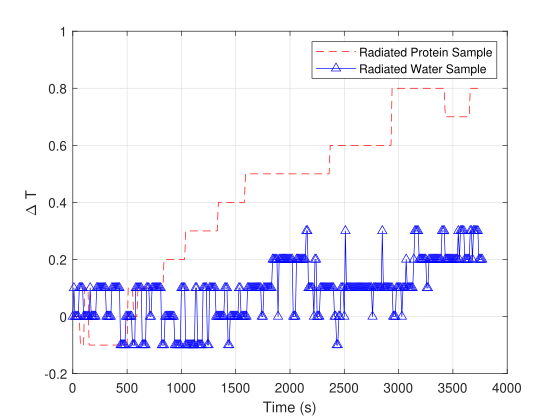}
\footnotesize
\caption{Temperature change versus time.}
\label{fig:temp_experiment}
\end{figure}

Fig.~\ref{fig:temp_experiment} presents the temperature change versus time experienced by both the radiated protein and water sample, respectively. A temperature reading was taken by the thermometer every 10 seconds. The experiment lasted for 1 hour. It can be seen that the radiated protein sample experiences a higher temperature change in comparison to the water sample. The purpose of the experiment is to show the impact of the radiated protein particles on the temperature change. Despite the fact that the egg-white is an aqueous solution, the availability of the protein particles results in the formation of a temperature gradient, which leads to a higher temperature increase in comparison to the water sample.

We emphasize that the main difference between the simulation results and the experimental setup is the frequency of operation. Given the availability of the sub-THz Tx operating at 130 GHz, it has been used as the radiation source. However, we believe that if we had a source with a frequency which matches the resonant frequency of the protein sample, a higher temperature increase would have been achieved at a shorter radiation time. Recall that in the case of the experiment, the Tx power is higher as the frequency of the device is lower. 

We also note that the experiment use a solution containing many protein types whereas our simulations considered only lysozyme. Overall, the conclusion attained from Fig.~\ref{fig:temp_experiment} overlaps with that given in Fig.~\ref{fig:temp_vs_power}, which indicates that at 130~GHz a lower temperature increase is attained in comparison to the 1.8~THz. This means that the simulation time should run longer to achieve an accumulative heating effect. 

Finally, the drop in temperature experienced at the beginning of the experiment can be attributed to the fact that the sample temperature tries to first achieve equilibrium with the surrounding temperature.

\section{Conclusions} 
\label{Sec:Sec6}
Protein molecules undergo resonance when exposed to THz frequencies that align with their vibrational modes. This results in the initiation of a thermal response due to their absorption of energy and converting it to heat. As such, proteins can be regarded as thermal nanosensors within an intra-body network. In this study, we investigate the influence of THz signals on protein heat dissipation. To achieve this, we employ a mathematical framework based on the heat diffusion model to examine how proteins absorb THz-EM energy and subsequently release it as heat into their surroundings. In addition, we conduct a parametric analysis to elucidate the effects of signal power, pulse duration, and inter-particle distance on protein thermal behavior.

Furthermore, we explore the correlation between temperature changes and the activation likelihood of thermally-gated ion channels. Our findings suggest that by selectively exciting protein particles at their resonant frequencies, precise temperature modulation can be attained within the intra-body environment. We support our findings through numerical simulations using COMSOL Multiphysics\textsuperscript{\textregistered} and propose an experimental setup to investigate the impact of THz radiation on protein behavior. This study concludes that controlled heating can lead to protein molecules acting as focal points that influence thermally-gated ion channels. Our work demonstrates the feasibility of engineering temperature changes at various temporal and spatial scales by manipulating THz-EM signals.

\bibliographystyle{IEEEtran}
\bibliography{IEEEabrv,./references}
 
\end{document}